 \documentclass[pmlr,twocolumn,10pt]{jmlr} 

\usepackage{booktabs}

\usepackage{graphicx}
\usepackage{booktabs}
 
\usepackage{siunitx}
\usepackage{amsmath,amssymb,amsfonts}
\usepackage{array,multirow}
\usepackage{mathtools}
\usepackage{dsfont}
\usepackage{pifont}
\usepackage{caption}
\usepackage{subcaption}
\usepackage{xcolor}  
\usepackage{tikz}
\usetikzlibrary{arrows, shapes, patterns}
\usepackage{adjustbox}

\usepackage[utf8]{inputenc} 
\usepackage[T1]{fontenc}    
\usepackage{booktabs}       
\usepackage{nicefrac}       
\usepackage{microtype}      
\usepackage{colortbl}
\usepackage[symbol]{footmisc}

\usepackage[switch]{lineno}

\newcommand{\equal}[1]{{\hypersetup{linkcolor=black}\thanks{#1}}}

\hypersetup{
    colorlinks=true,
    linkcolor=black,
    citecolor=blue,
    filecolor=black,
    urlcolor=black,
}

\theorembodyfont{\upshape}
\theoremheaderfont{\scshape}
\theorempostheader{:}
\theoremsep{\newline}

\jmlrvolume{259}
\jmlryear{2024}
\jmlrsubmitted{LEAVE UNSET}
\jmlrpublished{LEAVE UNSET}
\jmlrworkshop{Machine Learning for Health (ML4H) 2024} 

 \title[MpoxVLM]{MpoxVLM: A Vision-Language Model for Diagnosing Skin Lesions from Mpox Virus Infection}

\author{%
\Name{Xu Cao}\equal{Equal contribution} \Email{xucao2@illinois.edu}\\
\addr University of Illinois Urbana-Champaign, USA
\AND
\Name{Wenqian Ye}\footnotemark[1] \Email{wenqian@virginia.edu}\\
\addr University of Virginia, USA
\AND
\Name{Kenny Moise} \Email{kenny.moise@uniq.edu}\\
\addr Université Quisqueya, Haiti
\AND
\Name{Megan Coffee}\equal{Corresponding Author} \Email{megan.coffee@nyulangone.org}\\
\addr NYU Grossman School of Medicine, USA
}

\begin{document}

\maketitle

\begin{abstract}
In the aftermath of the COVID-19 pandemic and amid accelerating climate change, emerging infectious diseases, particularly those arising from zoonotic spillover, remain a global threat. Mpox (caused by the monkeypox virus) is a notable example of a zoonotic infection that often goes undiagnosed, especially as its rash progresses through stages, complicating detection across diverse populations with different presentations. In August 2024, the WHO Director-General declared the mpox outbreak a public health emergency of international concern for a second time. Despite the deployment of deep learning techniques for detecting diseases from skin lesion images, a robust and publicly accessible foundation model for mpox diagnosis is still lacking due to the unavailability of open-source mpox skin lesion images, multimodal clinical data, and specialized training pipelines. To address this gap, we propose MpoxVLM, a vision-language model (VLM) designed to detect mpox by analyzing both skin lesion images and patient clinical information. MpoxVLM integrates the CLIP visual encoder, an enhanced Vision Transformer (ViT) classifier for skin lesions, and LLaMA-2-7B models, pre-trained and fine-tuned on visual instruction-following question-answer pairs from our newly released mpox skin lesion dataset. Our work achieves 90.38\% accuracy for mpox detection, offering a promising pathway to improve early diagnostic accuracy in combating mpox.
\end{abstract}

\newcommand{\cmark}{\ding{51}}%
\newcommand{\xmark}{\ding{55}}%
\newcommand{\x}{\mathbf{x}}
\newcommand{\yhat}{\hat{y}_t}
\newcommand{\z}{\mathbf{z}}
\newcommand{\aaa}{\mathbf{a}}
\newcommand{\cc}{\mathbf{c}}
\newcommand{\p}{\mathbf{p}}
\newcommand{\ttt}{\mathbf{t}}
\newcommand{\mem}{\text{mem}}
\newcommand{\X}{\mathcal{X}}
\newcommand{\Y}{\mathcal{Y}}
\newcommand{\SSS}{\mathcal{S}}
\newcommand{\prlo}{p_\textit{lo}}
\newcommand{\prhi}{p_\textit{hi}}

\begin{keywords}
Vision-language models, Mpox, Infectious diseases, Skin Lesions
\end{keywords}

\paragraph*{Data and Code Availability}
We are using our own mpox dataset that we have built and collected, currently the largest of its kind globally, to conduct experiments for both baseline models and our proposed model. The dataset will be made available to other researchers upon submission of a formal request in accordance with IRB guidelines. To facilitate access, we will use a dedicated website to streamline the dataset sharing process. Additionally, the code used in our experiments will be released \href{https://github.com/IrohXu/MpoxVLM}{\underline{here}}


\noindent \textcolor{red}{\textbf{Content Warning:} This paper contains medical images that some may find sensitive.}

\section{Introduction}
\label{sec:intro}

Mpox is a zoonotic disease caused by the orthopoxvirus monkeypox virus (MPXV)~\citep{mitja2023mpox,mitja2023monkeypox,siegrist2023antivirals,laurenson2023description,lu2023mpox} that has affected over 99,518 individuals in an outbreak across 122 countries as of Sep 7, 2024. In the USA, there are 32,063 mpox cases with 58 associated deaths. Its spread across all inhabited continents has demonstrated sustained and prolonged human-to-human transmission, which was not previously recognized before 2022. The World Health Organization (WHO) declared its global spread a Public Health Emergency of International Concern (PHEIC) in 2022. Undetected infections delayed isolation and helped further transmission ~\citep{bragazzi2022knowing}. In July 2024, more than 100 laboratory-confirmed cases of clade 1b mpox were reported in South and North Kivu in the DRC and four neighboring countries — Burundi, Kenya, Rwanda, and Uganda— none of which had previously reported endemic mpox cases~\citep{rivers2024resurgence,sahin2022human}. The actual number of cases are expected to be higher, given case positivity and testing rates, as many clinically compatible cases have not been tested. Thus, it is very urgent to develop a computer-aided diagnosis method to detect mpox automatically, with high sensitivity for screening.

\begin{figure}[!t]
    \centering
    \includegraphics[width=\linewidth]{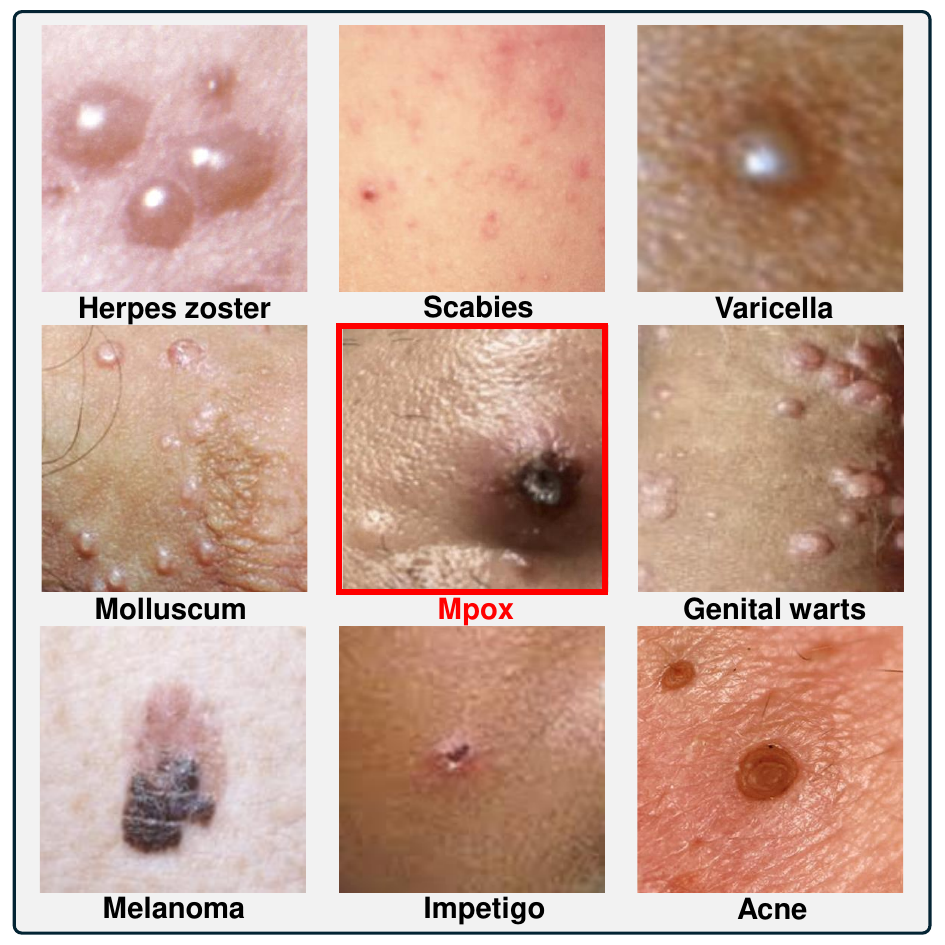}
    \caption{Examples of skin lesion images in our datasets. We have 86 classes (including mpox) in total. The medical images have been processed to remove any potential information leakage, such as hospital tags or disease information labels.}
    \label{fig:skin_example}
\end{figure}

Recently, the potential of deep learning, particularly through the use of convolutional neural network (CNN), has been explored to enhance mpox case detection using skin lesion images~\citep{khan2023recent, sizikova2022automatic,rampogu2023review,chadaga2023application,harris2024mpox,soe2024using}. However, vision models are susceptible to false negatives and errors, as relying on a single skin lesion image for disease detection can be highly inaccurate. In real-world clinical practice, medical professionals assess skin lesion images in conjunction with clinical, point of care and past laboratory values, epidemiological, demographic, and other relevant multimodal data to diagnose mpox. Additionally, many of the vision-based models developed for skin lesion analysis are trained on small, imbalanced dermatological datasets, often of low quality, including images of monkeys, drawings, and microscopic virus images~\citep{islam2022web}.

After the blooming of large language models (LLMs) and vision-language models (VLMs)~\citep{omiye2024large}, it becomes a wide trend to develop foundation models for different medical domains~\citep{li2023chatdoctor,huix2024natural}. Recent studies showed that VLMs finetuned with large-scale, broad-coverage biomedical figure-caption and visual question-answer datasets can effectively learn semantic characteristics via visual instruction tuning~\citep{liu2024visual,li2024llava}. Thus, connecting visual representations with multimodal text information from biomedical language models becomes increasingly critical to adapting foundation models for medical image classification, particularly in the challenging setting of multimodal data deficiency like mpox~\citep{cheng2023talk} and other skin lesion imaging tasks~\citep{zhou2023skingpt}.

In our study, we develop a VLM-based mpox diagnosis algorithm, MpoxVLM, designed specifically for the identification of mpox virus skin lesions in photographic skin lesion images together with easily accessible clinical information. Our algorithm's evaluation across our newly collected mpox dataset with multimodal information aims to ensure its effectiveness for individuals of different skin types, genders, ages, and lesion sites, thus facilitating timely and accurate diagnosis of mpox infections. The contributions are summarized as follows: 

\begin{enumerate}

    \item We propose the first Vision Language Model (VLM) framework MpoxVLM for mpox detection and its training pipeline. This is the first foundation model diagnosing skin lesions from mpox virus infection.
    
    \item We collected a new multimodal mpox diagnosis dataset from publicly available information including skin lesion images and clinical record from 2,914 samples.

    \item The experiment results on our newly collected dataset show that the proposed MpoxVLM achieves the best performance results than the previous state-of-the-art methods. 

\end{enumerate}

\section{Related Works}
\label{sec:rw}

\paragraph{Mpox History and How it spread.} 
Historically, the mpox was predominantly in children exposed to host animals, in outbreaks in Central and West Africa since the 1970s and later due to imported pets in the Midwestern United States in 2003, but the outbreak in Nigeria in 2017 and the later global outbreak since 2022 has been predominantly in adults~\citep{breman1980human,zinnah2024re,luna2022phylogenomic,guarner2022monkeypox}. The role of sexual networks was underappreciated until 2022 when transmission was largely among men who have sex with men (MSM)~\citep{allan2024prevalence,pekar2024genomic}. Given the changing epidemiology and the associated stigma and denial, cases, even in a known outbreak, were often missed; in one study, only 23\% of healthcare workers used appropriate personal protective equipment for all encounters with patients later confirmed to have mpox~\citep{marshall2022health}. Mpox is often mistaken for another infection, such as sexually transmitted infections and viral exanthems, but also less common autoimmune and infectious causes. The lesions develop from unremarkable pimples over days to weeks into pathognomonic, umbilicated pustular lesions, but are infectious throughout and transmission often occurs before diagnosis is suspected. Associated symptoms (fever, sore throat, lymphadenopathy, and pain) vary between patients and resemble other diseases, so clinical suspicion usually focuses on the rash. Given a missed infection can lead to the further community - or less commonly, healthcare clinic transmission - it is important to have a screening test and clinical support for clinicians with little experience diagnosing mpox. Thus, it is very meaningful to develop and deploy computer-aided diagnosis solutions for those countries influenced by mpox~\citep{bleichrodt2024evaluating}.

\paragraph{Mpox Diagnosis with Deep Learning.} 
Deep learning has shown promise in classifying skin lesions in dermatology~\citep{barata2019deep,liu2020deep,groger2023towards}. For mpox, Convolutional Neural Networks (CNNs) have demonstrated effectiveness in identifying disease through the analysis of skin lesion images~\citep{thieme2023deep}. Initial studies employed CNN models such as MobileNet~\citep{jaradat2023automated,altun2023monkeypox}, VGG Net~\citep{ahsan2022image} that were pre-trained using the ImageNet database, and subsequently fine-tuned with either publicly or privately sourced mpox skin lesion image datasets. Sitaula et al.~\citep{sitaula2022monkeypox} further proposed an ensemble learning method to combine Xception and DenseNet to predict mpox. Bala et al.~\citep{bala2023monkeynet} amassed a dataset comprising 770 skin lesion images, including 279 mpox cases, and introduced MonkeyNet, a model that leverages the architecture of DenseNet alongside multiple data augmentation techniques. Subsequent research has made the Mpox Skin Lesion Dataset (MSLDv2.0)~\citep{ali2023web} publicly accessible, which includes 284 Mpox image samples. However, these datasets are still too small to train reliable models.

\paragraph{Vision-Language Models (VLMs).}
In computer vision, the exploration of VLMs for handling tasks involving multiple modes of data has developed very fast. This interest has sparked new developments in multi-modal large language models (MLLMs), leading to the creation of new innovative tools like GPT-4V~\citep{wu2023can} and Gemini~\citep{team2023gemini}. A key strategy to embed visual information into language models involves the fine-tuning or instruction tuning of large-scale foundational VLMs, such as Flamingo~\citep{alayrac2022flamingo} and MiniGPT4~\citep{zhu2023minigpt}. This technique has been further refined by recent developments like visual prompt tuning~\citep{jia2022visual}, LLaVA~\citep{liu2024visual,li2024llava}, InternVL~\cite{chen2024internvl}, LLaMA-Adapter~\citep{zhang2023llama}, and BLIP-2\citep{li2023blip}, which employ instruction-following LLMs trained on VQA datasets specifically designed for image instruction tuning. The effectiveness of instruction tuning in improving the performance of VLMs on multi-modal vision-language tasks has been demonstrated, highlighting its potential to significantly advance the field.

\section{Methodology}
\label{sec:method}

\paragraph{Overall Architecture.} Fig~\ref{fig:framework} illustrates the main workflow of the proposed MpoxVLM framework. It included two encoders and a LLM decoder. The first encoder is a Contrastive Language-Image Pretraining (CLIP) visual encoder and it is frozen during our model training. The second encoder is a Vision Transformer (ViT) classifier encoder self-supervised learning pretrained with our proposed mpox dataset in using masked autoencoder~\citep{he2022masked}. The LLM decoder used in our framework is LLaMA-2-7B~\cite{touvron2023llama} with 7 billion parameters. Each encoder is connected with the LLM decoder by a two-layer multilayer perceptron (MLP) module as the alignment between visual features and language features.

Inspired by LLaVA~\citep{liu2023improved} and MiniGPT-4~\citep{zhu2023minigpt}, our framework aims to integrate the visual features of skin lesion images and the clinical records of patients into a unified representation, leveraging the LLM to facilitate the diagnosis of mpox. The input of the task comprises a skin lesion image, denoted as $X_v$, alongside context-specific prior knowledge, represented as $X_c$. 

\begin{figure*}[!htbp]
    \centering
    \includegraphics[width=1.00\linewidth]{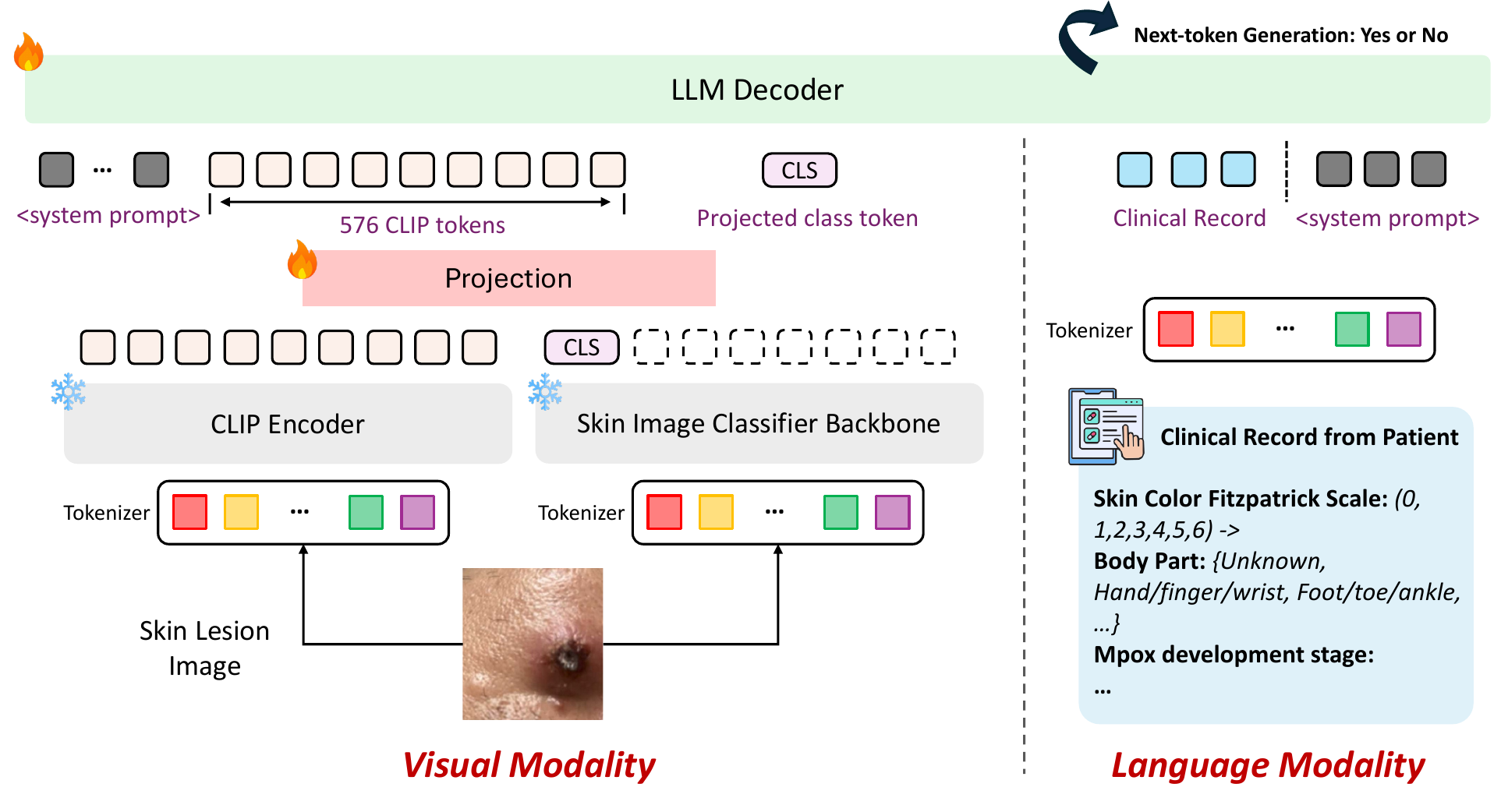}
    \caption{Workflow of the MpoxVLM framework and visual question-answer pair examples.}
    \label{fig:framework}
    \vspace{-0.2cm}
\end{figure*}

To train the MpoxVLM, we also design question-answer pair from our proposed dataset, $(X_q, X_o, Y)$, to guide the model's learning. The aim of using question-answer pairs as input and output structures is to streamline the integration of visual and textual data. Here, $X_q$ poses a critical question based on the evaluation of the skin lesion image: "After reviewing this skin lesion image, do you think the patient has mpox?". $X_o$ is a group of options denoting mpox or non-mpox. In our original dataset, $X_o$ can be multiple choices including $>30$ common skin diseases. In order to let model only focus on the diagnosis of mpox, we set the number of options to two in the training phase. The task of MpoxVLM and other baseline models is to predict an answer $\hat Y$ from the options, which represents the model's assessment of whether the patient is likely to have mpox. 

The proposed MpoxVLM $F_{\theta}$ can be formulated as:

\begin{equation}
    \hat Y = F_{\theta}(X_v, X_c, X_{q}, X_{o})
\end{equation}

In the following subsections, we will introduce the structure of MpoxVLM and our training procedure.

\subsection{Model Design}

\paragraph{Visual Encoder.} 
Different to previous VLM solutions in medical imaging such as Med-LLaVA, the MpoxVLM framework incorporates dual encoders to process and analyze input image data effectively. The first encoder $f_{\operatorname{CLIP}}$ is a visual-language encoder, specifically a freezed pre-trained CLIP visual encoder. Within the MpoxVLM framework, this encoder's role is to extract general visual features from the input mpox image, enabling a detailed visual-language representation of the skin lesion image. The extracted features are then passed through a trainable projection alignment module (a 2-layer MLP). This alignment module is crucial for mapping the visual tokens into text embedding space of the LLM. The second encoder $f_v$ in the MpoxVLM system is a visual encoder, built upon a pre-trained vision transformer (ViT) classifier that is specialized in skin lesion image analysis. This encoder focuses on deriving high-level classification insights from the visual data. The output from this visual encoder, particularly the classification token, is directed through another trainable projection alignment module. The purpose of this alignment module is to project the classification token accurately into the LLM's text embedding space. These dual encoders are important for translating complex visual features into a format that the LLM can seamlessly process and interpret, enhancing the overall visual understanding capability of the MpoxVLM.

\paragraph{Visual Instruction Tuning for MpoxVLM.} To finetune the LLM within the MpoxVLM framework, our approach aims to synergistically harness the strengths of the pretrained LLM (LLaMA-2) along with the integrated dual visual encoders.  For input skin lesion image $X_v$, we use the tokenizer from CLIP visual encoder (ViT-L/14-336) to embed the image into tokens. The visual feature is extracted from CLIP's visual encoder and then the adapter layer to map image features into the LLM's word embedding space:

\begin{equation}
    \begin{split}
    Z_{\operatorname{CLIP}} = W_{\operatorname{CLIP}} \cdot f_{\operatorname{CLIP}}(\operatorname{Tokenizer}(X_{v})),\;\;\; \\ 
    Z_{\operatorname{CLIP}} \in R^{d_h \times k}
    \end{split}
\end{equation}

, where $f_{\operatorname{CLIP}}$ is the CLIP encoder. $W_{\operatorname{CLIP}}$ is the weight of the linear projection layer. $d_h$ is the dimension of the LLM embedding. $k$ is the number of visual tokens.

In the classification path, the input skin lesion image $X_v$ is tokenized by the same CLIP's patch embedding layer and then extracted visual tokens by a pretrained ViT for skin lesion image classification. The classification token of the last layer is then fed to an adapter layer to map the classification features into the LLM's word embedding space.

\begin{equation}
    \begin{split}
    Z_{\operatorname{V}} = W_{\operatorname{V}} \cdot [f_{\operatorname{V}}(\operatorname{Tokenizer}(X_{v}))]_{\operatorname{CLS}},\;\;\; \\ 
    Z_{\operatorname{V}} \in R^{d_h \times 1}
    \end{split}
\end{equation}

, where $f_{\operatorname{V}}$ is the ViT encoder for classification. $W_{\operatorname{V}}$ is the weight of the linear projection layer. $d_h$ is the dimension of the LLM embedding. $\operatorname{CLS}$ means the classification token from the ViT's output.

For a sequence of length $L$, the autoregressive encoder in the LLM for generation mpox diagnosis answer is as follows:

\begin{equation}
\begin{aligned}
    &p(\hat Y | Z_{\operatorname{CLIP}}, Z_{\operatorname{V}}, X_c, X_{q}) \\
    & = \prod_{i=1}^{L} p_{\theta}(y_i | (Z_{\operatorname{CLIP}} \oplus Z_{\operatorname{V}}), X_c, X_{q}, X_{o}, \hat Y_{<i})
\end{aligned}
\end{equation}

, where $X_{q}$ is all of the question tokens (the whole question). $X_{o}$ is the option tokens. $\oplus$ is the concatenation operation for latent features. $\hat Y_{<i}$ is all answer tokens before $y_i$.

\begin{equation}
    \hat{Y} = f_{\operatorname{LLM}}(Z_{\operatorname{CLIP}}, Z_{\operatorname{V}}, X_c, X_{q}, X_{o})
\end{equation}

, where $f_{\operatorname{LLM}}$ is the LLM (LLaMA-2-7B).


\subsection{Training Pipeline}

Firstly, we pretrained a ViT encoder (ViT-L-14-336) with the proposed mpox skin lesion dataset. Then, we build the MpoxVLM model with the pretrained ViT encoder as the classification visual encoder and the CLIP encoder (clip-vit-large-patch14)~\citep{radford2021learning} as the visual-language encoder. Then, in the following experiment, we freeze the weights of both the CLIP visual encoder and classification visual encoder. The LLM used in our task is LLaMA-2 7B~\citep{touvron2023llama}. To train the two linear adaptors, we keep LLM weights frozen and only train alignment layers between the dual visual encoders and the LLM. After pretraining the alignment layers, we keep all weights frozen except the weight of LLMs, and finetune the weight under the mpox visual question answer task.

\section{Experiments and Results}
\label{sec:experiment}

\begin{figure*}[h]
    \centering
    \includegraphics[width=\linewidth]{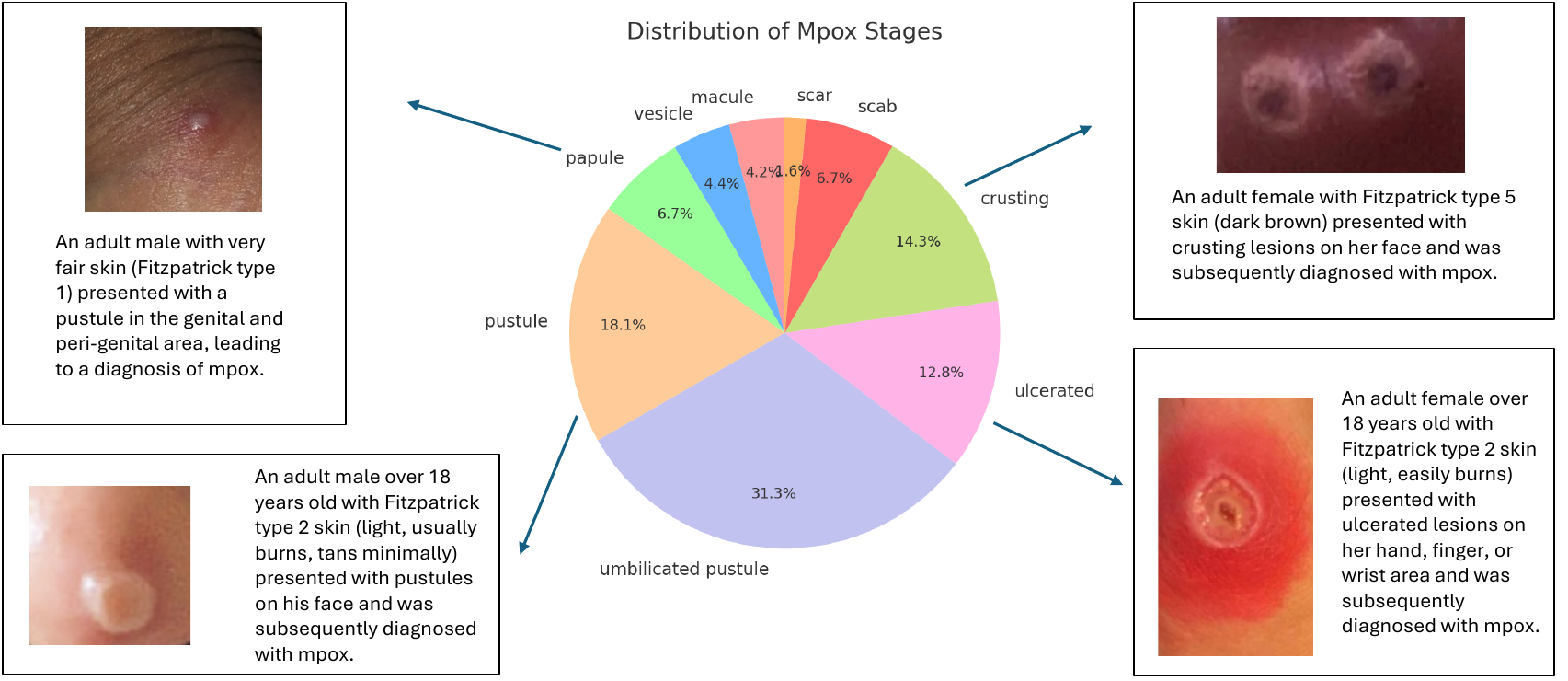}
    \caption{Distribution of Mpox development stages and instruction tuning data in the mpox dataset. The stages of Mpox progression are ordered as follows: macule, vesicle, papule, pustule, umbilicated pustule, ulceration, crusting, scab, and scar.}
    \label{fig:vis}
\end{figure*}


\subsection{New Mpox Dataset and Demographic Analysis.}

\begin{figure}[h]
    \centering
    \includegraphics[width=0.9\linewidth]{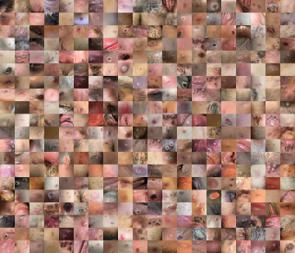}
    \caption{Mpox skin images in our dataset.}
    \label{fig:framework}
\end{figure}

Prior to our work, there are some mpox skin lesion image datasets available~\citep{ahsan2022monkeypox,jaradat2023automated,bala2023monkeynet,ali2023web}. However, most of the datasets are relatively small and can not be used to train large models. To solve this issue, we proposed a new mpox dataset including both mpox skin lesion image and multimodal clinical information such as patients' medical history. The comparison of different mpox datasets is presented in Table~\ref{tab:data}. All of the data in our new dataset is collected and curated by doctors from public sources (medical and news journals, public health websites, and social media). It comprises 1,057 Mpox-positive samples and 1,857 samples of other diseases, totaling 2,914 samples. We have meticulously cropped each image to remove surrounding clothing or backgrounds and to focus on disease lesions and supplemented them with demographic annotations in Figure~\ref{fig:vis} for skin type (Fitzpatrick scale of 1-6), gender or sex presentation (based on accompanying information or image alone), and age (adult or child). The dataset also included the affected body part and stage of mpox lesions (macule, papule, vesicle, pustule, umbilicated pustule, ulcerating, crusting, scab, to scar) in order to ensure early stages were represented, as many images are of late, classic stages. Doctors carefully selected comparator images from 86 diagnoses that closely resemble mpox. These images were chosen to match key characteristics, including demographic factors (skin type, gender, age), affected body parts (as initial lesions often correspond to the transmission route and area of first exposure), and the epidemiology of diseases commonly mistaken for mpox. Additionally, some diseases that clinicians may be less familiar with were included, recognizing these may be more difficult to distinguish. Infectious disease specialists selected the comparator diseases based on global epidemiology and clinical presentations, using both clinical physician expertise and image searches to find visual mimics. The primary objective of this dataset refinement is to fully leverage the data-centric approach of large foundation models. This approach is aimed at achieving a more accurate representation of data distribution and facilitating a fairer comparison with previously established methodologies.

\begin{figure}[!htbp]
    \centering
    \includegraphics[width=\linewidth]{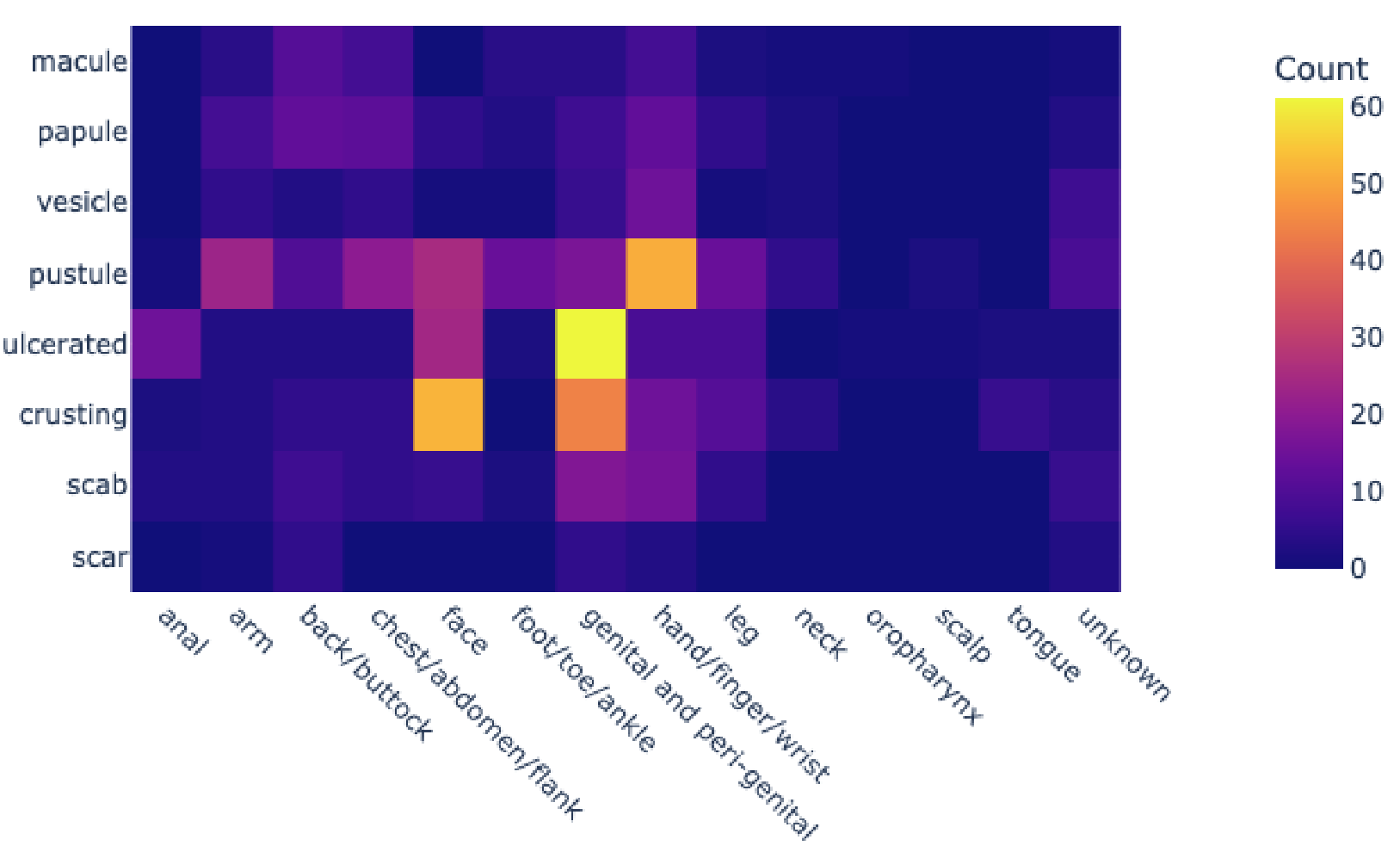}
    \caption{Heatmap illustrating the distribution of mpox across various body parts (x-axis) and stages of the disease (y-axis). The dataset covers nearly all body parts where mpox lesions are likely to occur. Although occurrences on areas such as the tongue and neck are very rare, these regions are still included in the dataset for completeness.}
\end{figure}

\begin{table*}[!htbp]\small
\centering
\begin{tabular}{lccc}
\toprule
Dataset  & Mpox Sample Size & All Sample Size  & Demographic Information  \\
\midrule
Ahsan et al.~\citep{ahsan2022monkeypox}$^{*}$ & 43 & 161 &  \xmark  \\
Jaradat et al.~\citep{jaradat2023automated}$^{*}$ & 45 & 117 &  \xmark  \\
MonkeyNet~\citep{bala2023monkeynet} & 178 & 492 &  \xmark  \\
MSLD v2.0~\citep{ali2023web} & 284 & 755 &  \cmark  \\
Kularathne et al.~\citep{kularathne2024mpox} & 120 & 120 &  \xmark  \\
\midrule
\textbf{Ours} & \textbf{1,057} &  \textbf{2,914}  & \cmark  \\
\bottomrule
\end{tabular}
\caption{Mpox Dataset Comparison. * denote original data (without data augmentation). Kularathne et al.~\citep{kularathne2024mpox} used 120 data to synthesize 1,000 mpox images via DreamBooth~\citep{ruiz2023dreambooth}.}
\label{tab:data}
\end{table*}

\begin{table*}[!htbp] \small
\centering
\label{tab:performance}
\begin{tabular}{lcccccc}
\toprule
Methods  &  Accuracy$(\%)\uparrow$  &  Precision$(\%)\uparrow$  &  Recall$(\%)\uparrow$  &  F1 Score$(\%)\uparrow$  &  AUROC$(\%)\uparrow$ \\
\midrule
\citet{bala2023monkeynet} & $73.06_{\pm1.79}$  & $66.45_{\pm 1.37}$ & $54.21_{\pm4.21}$ & $59.71_{\pm3.63}$ &  $78.23_{\pm3.04}$ \\ 
\citet{thieme2023deep}  & $74.61_{\pm0.78}$ & $66.48_{\pm3.33}$ & $62.63_{\pm4.21}$ & $64.50_{\pm0.89}$ & $80.88_{\pm0.61}$  \\ 
\citet{sitaula2022monkeypox} & $87.21_{\pm1.74}$ & $85.62_{\pm0.08}$ & $78.42_{\pm5.79}$ & $81.87_{\pm3.24}$ & $93.91_{\pm0.66}$  \\ 
\citet{ahsan2022image} & $74.42_{\pm1.63}$ & $77.36_{\pm1.24}$ & $43.16_{\pm4.25}$ & $55.41_{\pm3.71}$ & $79.55_{\pm2.94}$  \\
\citet{jaradat2023automated} & $81.20_{\pm1.41}$ & $71.43_{\pm1.56}$ & $81.58_{\pm2.11}$ & $76.17_{\pm1.81}$ & $89.33_{\pm0.80}$  \\
\citet{altun2023monkeypox} & $79.46_{\pm0.31}$ & $74.14_{\pm1.26}$ & $67.89_{\pm1.97}$ & $70.88_{\pm0.66}$ & $83.38_{\pm0.91}$  \\
\citet{li2024llava} & $87.34_{\pm0.55}$ & $84.16_{\pm1.89}$ & $82.33_{\pm2.51}$ & $82.24_{\pm1.22}$ & $85.95_{\pm0.73}$ \\
\midrule
\textbf{Ours} & $\textbf{90.38}_{\pm0.80}$ & $\textbf{86.77}_{\pm1.85}$ & $\textbf{84.38}_{\pm0.78}$ & $\textbf{85.72}_{\pm1.14}$ & $\textbf{95.16}_{\pm0.72}$  \\
\bottomrule
\end{tabular}
\caption{Comparisons with previous methods. For all mpox skin lesion image detection baseline models, we finetuned there model in our proposed dataset, including the new open-source medical VLM Med-LLaVA.}
\end{table*}

\subsection{Evaluation Metrics.} To evaluate the performance of our model, we use several standard metrics: \textbf{Accuracy:} measures the proportion of correctly predicted instances out of all predictions. \textbf{Precision:} known as \textit{positive predictive value}. It evaluates the proportion of true positive predictions to the total number of positive predictions. It is useful for understanding the reliability of positive classifications made by the model. \textbf{Recall:} known as \textit{sensitivity}, it assesses the proportion of actual positives correctly identified by the model, which is crucial in medicine for designing a screening test. \textbf{F1 Score:} the harmonic mean of precision and recall. It considers both false positives and false negatives. \textbf{AUROC:} The Area Under the Receiver Operating Characteristic curve represents the likelihood of the model distinguishing between classes. A higher AUROC score indicates better model performance.

\subsection{Training Details.}
As most prior papers did not provide open-sourced code, therefore, we reproduce their methods carefully and trained them on our dataset. We adopt a dataset split of \textbf{5:1:1 (2000:398:398)} for training, validation, and testing set, respectively. The split was done randomly based on patients' ID. The ratio of positive (mpox) and negative (non-mpox) is 4:7 in the test set.

The training setting for all baselines is the same. To preprocess the images, we also center-crop and resize the input images to $336\times 336$ resolution. The AdamW optimizer~\citep{loshchilov2017decoupled} with a weight decay set at 0.01 is also used to train all baselines and MpoxVLM. Additionally, we employ a cosine annealing learning rate scheduler~\citep{loshchilov2016sgdr}, with the base learning rate set at $5\times 10^{-5}$. All baseline models are trained for 100 epochs and use early stops. The MpoxVLM is finetuned for 10,000 steps on 2 NVIDIA RTX 4090 GPUs using LoRA~\citep{hu2021lora}, with each GPU handling a batch size of 2. All codes and dataset details will be available upon acceptance.

\subsection{Compare with Previous Methods.}
This section provides a comparative analysis of our method against existing approaches in the field.  Previous studies~\citep{bala2023monkeynet,thieme2023deep,ahsan2022image,jaradat2023automated,altun2023monkeypox} have predominantly employed CNN-based single network architectures. Our method demonstrates a significant improvement in performance over these CNN-based approaches. Conversely, another line of research, exemplified by Sitaula et al.~\citep{sitaula2022monkeypox}, utilizes an ensemble of multiple networks. Different from these methods, our method adopts the idea that LLM can enhance the performance of ViT in classification tasks~\citep{chu2024visionllama} and uniquely leverages visual-language features extracted with general domain to enhance specific domain knowledge in the mpox diagnosis area. 

\subsection{Compare with Multimodal LLMs.}

As multimodal LLMs have been applied to various medical imaging tasks~\citep{wu2023can,panagoulias2024evaluating} and are proved having basic understanding of mpox~\citep{cheng2023talk}, we also evaluated their ability to diagnose skin lesions caused by mpox virus infection. The input to these models included a system prompt: ``Mpox is a zoonotic disease caused by the orthopoxvirus monkeypox virus. The example image is a patient with mpox. You are a helpful visual reasoning assistant that can distinguish mpox in new skin lesion images," along with an example mpox skin image. The experimental results, shown in Table~\ref{tab:mllms}, reveal that state-of-the-art multimodal LLMs, such as Claude 3 variants~\citep{claude3_family} and GPT-4o~\citep{achiam2023gpt}, perform only marginally better than random guessing. Despite advancements, these models still struggle with fine-grained clinical tasks, suggesting that in their current unprimed state, they are not yet suitable for complex clinical applications. In contrast, the MpoxVLM framework achieves a significantly higher accuracy of 90.38\%, underscoring the importance of our new mpox skin image dataset in fine-tuning VLMs for precise mpox diagnosis.

\subsection{Ablation Study.} To validate the effectiveness of our model design, we conduct a detailed analysis of the individual contributions of various modules within our framework. We observe that even by freezing the weight of the CLIP model and only fine-tuning the LLM (LLaMA-2 7B), the performance can achieve relatively high accuracy similar to the previous SOTA method (line 2). Another key observation is the significant impact of the classification visual encoder on our framework's performance. This is obvious in the marked improvement in the AUROC metric.

\begin{table}[t]
    \centering
    \adjustbox{max width=\linewidth}{
    \begin{tabular}{cc}
    \toprule
    Multimodal LLMs & Accuracy$(\%)\uparrow$  \\ 
    \hline
       Claude 3 Haiku  & 53.07\% \\
       Claude 3 Sonnet  & 52.98\%  \\
       Claude 3 Opus  & 52.48\% \\
       Claude 3.5 Sonnet  & 62.51\%  \\
       GPT-4o-mini & 67.22\%  \\
       GPT-4o  & 76.45\%  \\
    \hline
       MpoxVLM (Ours)  & \textbf{90.38}\%  \\
    \bottomrule
    \end{tabular}
    }
    \caption{Comparison with state-of-the-art Multimodal LLMs. Note: To let these closed source models answer the mpox skin lesion image questions, we use the same system prompt to provide LLMs related clinical background to mpox and use in-context learning to guide these models to answer the question.}
    \label{tab:mllms}
\end{table}

\begin{table}[t]
    \centering
    \adjustbox{max width=\linewidth}{
    \begin{tabular}{cccc|cc}
        \toprule
        Classifier & CLIP & Text & LLMs & Accuracy$(\%)\uparrow$  & AUROC$(\%)\uparrow$ \\
        \midrule
        \cmark & \xmark & \xmark & \xmark & $87.94_{\pm0.74}$ & $94.88_{\pm0.88}$ \\
        \xmark & \cmark & \xmark & \cmark & $87.34_{\pm0.55}$ & $85.95_{\pm0.73}$ \\
        \xmark & \cmark & \cmark & \cmark &  $88.32_{\pm0.12}$ & $88.44_{\pm0.16}$\\
        \cmark & \cmark & \cmark & \cmark & $\textbf{90.38}_{\pm0.80}$ & $\textbf{95.16}_{\pm0.72}$ \\ 
        \bottomrule
    \end{tabular}
    }
    \caption{Ablation study on different modules. The first row (Classifier only) refers to using only the Vision Transformer classifier for the task (similar to \citep{kularathne2024mpox}). The second row (CLIP+LLMs) represents finetuning the LLaVA-based VLM baseline without incorporating additional Vision Transformer classifier module, with the model’s weights initialized by LLaVA-Med~\citep{li2024llava}. The third row (CLIP+Text+LLMs) refers to finetuning the VLM baseline while incorporating additional clinical history and demographic information about the patients. The fourth row showcases the results from MpoxVLM.}
    \label{tab:ab}
\end{table}


\section{Discussion}
\label{sec:discussion}

Mpox exemplifies the domain gap issue expected in emerging diseases. Tools trained on initial, small datasets do not capture the diversity in a mature epidemic. Here an outbreak in a specific population spreads to a broader population requiring training of images of different lesion stages on a broad range of skin types and body parts. To ensure this, we use a significantly larger dataset than prior mpox computer vision approaches. This dataset encompasses a diverse demographic, mirroring those affected by mpox - as the epidemiology shifted from children in West and Central Africa to the Midwestern United States, to adults in Nigeria, and more broadly in MSM globally, with people of color disproportionately affected. The current mpox outbreak with new variant clade Ib has been affecting both adults and children in Burundi and the DRC, but with lower case positivity rates in children tested, we need to ensure accuracy in diagnosis, given the higher mortality rates of mpox in young children, but also the risks of isolating a young child and of not treating other causes of rash and illness. Images also reflect the range of mpox stages; many available images reflect the more easily identified, later images (such as the crusting umbilicated pustules in Fig 1) but fail to include earlier, less recognizable stages such as papules, whose identification would reduce transmission. Comparator diseases reflect both demographics and the global disease landscape against which mpox is identified. In addition, the images focus on the lesion alone, to avoid spurious correlations with clothing, backdrops, or even image quality, which may hint at disease geographies. 

The scarcity of mpox images in the existing literature underscores the neglect of a disease known for 50 years to be endemic in West and Central Africa. In 2023, at least 581 people, mostly young children, likely died from mpox, with the case-fatality ratio attaining 8\% in highly affected areas~\citep{who2024mpox}. However, many past dermatologic tools have included few images of persons of color or, as seen here, do not describe the demographics of the dataset. Such a tool will also need to be practical and achieve high accuracy, here 90\%, and a high recall (sensitivity) for screening, and precision (positive predictive value) to assist in diagnosis where laboratory capacity is limited. Given the expected rise in novel and emerging diseases due to climate change and increased global movement, such an approach leads the way for further tools to address novel diagnostic challenges~\citep{baker2022climate}.

\section{Conclusion}
\label{sec:conclusion}
Mpox has been historically neglected - first in Central and West Africa - and then in vulnerable populations globally including the LGBTQ community and disproportionately among people of color. The lack of attention to mpox has resulted in gaps in research and resources, particularly in dermatologic imaging and diagnosis. This paper outlined the first application of a vision-language model specifically tailored for dermatologic mpox images, alongside the largest fine-grained diagnostic dataset/benchmark for skin lesions caused by mpox. Our framework integrates clinical information such as mpox stages, affected body parts, gender, age, and Fitzpatrick skin types, making it the most comprehensive dataset for both AI researchers and clinicians. By developing this framework, we aim to provide a critical tool that supports clinicians in diagnosing mpox and, in the future, other emerging infectious diseases from skin lesion images. In future work, we aim to publish larger datasets containing more samples for each class and design the vision foundation model for digital dermatology.

\bibliography{references}
\clearpage
\appendix

\begin{table}[!htb]
\centering
\caption{Skin Color Fitzpatrick Scale}
\label{tab:skin}
\begin{tabular}{|c|l|}
\hline
\textbf{Number} & \textbf{Description} \\
\hline
0 & Unknown \\
1 & Always burns, never tans (palest; freckles) \\
2 & Usually burns, tans minimally (light colored but darker than fair) \\
3 & Sometimes mild burn, tans uniformly (golden honey or olive) \\
4 & Burns minimally, always tans well (moderate brown) \\
5 & Very rarely burns, tans very easily (dark brown) \\
6 & Never burns (deeply pigmented dark brown to darkest brown) \\
\hline
\end{tabular}
\end{table}

\section{Description of Dataset Terms}

This appendix provides a detailed explanation of the terms used in our dataset, explaining the meaning of the numbers corresponding to different attributes.
Table ~\ref{tab:skin} describes the Fitzpatrick Skin Type Scale, which categorizes skin types based on their response to sun exposure. The numerical values in this table range from 0 (Unknown) to 6, with each number representing a specific skin type. The scale is widely used in dermatology to classify different levels of melanin in the skin, influencing susceptibility to sunburn and skin cancer risk.

Table ~\ref{tab:body} provides the body parts where mpox lesions can be observed. Each body part is assigned a numerical value from 0 (Unknown) to 11, with each number corresponding to a specific region of the body.

Table ~\ref{tab:disease} lists the different diseases or conditions in the dataset and assigns each a corresponding numerical value. The numbers range from 0 (Unknown disease) to 86, covering a wide range of skin-related infections, inflammatory conditions, and benign or malignant skin lesions. This table is important for annotating and diagnosing different dermatologic conditions alongside mpox.

\begin{table}[!htb]
\centering
\caption{Body Part}
\label{tab:body}
\begin{tabular}{|c|l|}
\hline
\textbf{Number} & \textbf{Body Part} \\
\hline
0 & Unknown \\
1 & Hand/finger/wrist \\
2 & Foot/toe/ankle \\
3 & Arm \\
4 & Leg \\
5 & Chest/Abdomen/flank \\
6 & Back/Buttock \\
7 & Face \\
8 & Neck \\
9 & Genital and peri-genital \\
10 & Anal \\
11 & Scalp \\
\hline
\end{tabular}
\end{table}

\begin{table*}[htb]
\centering
\caption{Disease in the data.}
\label{tab:disease}
\begin{tabular}{|c|l|c|l|}
\hline
\textbf{Number} & \textbf{Disease} & \textbf{Number} & \textbf{Disease} \\
\hline
0 & Unknown & 43 & Other folliculitis \\
1 & Varicella (chickenpox) & 44 & Other mycobacterial infections \\
2 & Herpes zoster (shingles) & 45 & Bartonella \\
3 & Measles & 46 & PLEVA \\
4 & Herpes - extra-genital & 47 & Other rickettsia or scrub typhus \\
5 & Herpes - genital & 48 & Melioidosis \\
6 & Syphilis primary or congenital & 49 & Lichen Planus \\
7 & Syphilis secondary & 50 & Buruli Ulcer \\
8 & Erythema Migrans & 51 & Talaromyces marneffei \\
9 & Healed Scar & 52 & Coccidioidomycosis \\
10 & Acne & 53 & Paracoccidioidomycosis \\
11 & Molluscum & 54 & Sporotrichosis \\
12 & Scabies & 55 & Fusarium \\
13 & Hives, urticaria & 56 & Leukemic Cutis \\
14 & Skin cancer & 57 & Eczema \\
15 & Tularemia & 58 & Roseola \\
16 & Blastomycosis & 59 & Perforating Dermatoses \\
17 & Hand foot and mouth disease & 60 & Disseminated Gonorrhea \\
18 & Impetigo or ecthyma & 61 & Dermatitis herpetiformis \\
19 & Bed bug bites & 62 & Prurigo Nodularis \\
20 & Other insect bites & 63 & Nipple disorder \\
21 & Genital warts (HPV) & 64 & Drug Eruption \\
22 & Furunculosis or early abscesses & 65 & Bullous Pemphigoid \\
23 & Folliculitis (standard bacterial) & 66 & Dermatofibroma \\
24 & Miliaria & 67 & Plain skin (no disease) \\
25 & Tuberculosis & 68 & Nipple (no disease) \\
26 & BCG vaccination & 69 & Genital skin (no disease) \\
27 & Lymphangioma circumscriptum & 70 & Anal area (no disease) \\
28 & Spider bite & 71 & Oral lips (no disease) \\
29 & Herpes gestationis & 72 & Nose (no disease) \\
30 & Donovanosis & 73 & Scalp with hair (no disease) \\
31 & Behcet's & 74 & Beard (no disease) \\
32 & Leishmaniasis & 75 & Hemorrhoid \\
33 & Chancroid & 76 & Anal (no disease) \\
34 & Erythema multiforme & 77 & Freckled skin (no disease) \\
35 & Toxoplasmosis & 78 & Mole (no disease) \\
36 & Histoplasmosis & 79 & Pyogenic granuloma \\
37 & Rickettsia Akari & 80 & Janeway lesions \\
38 & Cryptococcus & 81 & Palm (no disease) \\
39 & Degos & 82 & Buttock (no disease) \\
40 & Rickettsia parkeri & 83 & Teeth (no disease) \\
41 & Pityriasis rosea & 84 & Teeth (caries) \\
42 & Psoriasis (guttate) & 85 & Anthrax \\
 &  & 86 & Malakoplakia \\
\hline
\end{tabular}
\end{table*}
\end{document}